\documentclass[review,12pt]{elsarticle}

\usepackage{showlabels}
\usepackage[colorlinks=true, pdfstartview=FitV, linkcolor=red, citecolor=blue, urlcolor=blue]{hyperref}
\usepackage{bbm}
\usepackage{graphicx}
\usepackage{amsmath}
\usepackage{amsfonts,amsbsy}
\usepackage{amssymb}

\newcommand{\beq}{\begin{equation}}
\newcommand{\eeq}{\end{equation}}
\newcommand{\bea}{\begin{eqnarray}}
\newcommand{\eea}{\end{eqnarray}}

\usepackage{mathtools}

\newcommand\myeq{\stackrel{\mathclap{\normalfont\mbox{Large $N_c$}}}{=}}

\begin{document}

%\preprint{ }

\title{Finite Numbers of Sources, Particle Correlations and  the Color Glass Condensate}
\author[bnl,ccnu]{Larry McLerran}
\author[rbrc]{Vladimir V. Skokov}
%\email{mclerran@bnl.gov}
\address[bnl]{
Department of Physics, Brookhaven National Laboratory, 
Upton, NY 11973}
\address[rbrc]{RIKEN/BNL, Brookhaven National Laboratory, 
Upton, NY 11973}
\address[ccnu]{Physics Dept, China Central Normal University, Wuhan, China}

%\email{vskokov@quark.phy.bnl.gov}

\begin{abstract}
We show that for a finite number of emitting sources, the Color Glass Condensate produces substantial elliptic azimuthal anisotropy, characterized by  $v_2$, for two and four particle correlations for momentum greater than or of the order the saturation momentum.  The flow produced has the correct semi-quantitative features to describe flow seen in LHC experiments with p-Pb and pp collisions. This flow is induced by quantum mechanical interference between the waves of produced particles, and the flow itself is coupled to fluctuations in the positions of emitting sources. We shortly discuss
 generalizing these results to odd $v_n$, to correlations involving larger number of particles, and to transverse  momentum scales $\Lambda_{\rm QCD} \ll p_T \ll  Q_{\rm sat}$.
\end{abstract}

\maketitle

\section{Introduction}

It has been shown in the LHC experiments that in both $pA$ and $pp$ collisions, 
there is a substantial azimuthal anisotropy, quantified by the second Fourier harmonics also often referred to as elliptic ``flow'',~\cite{Khachatryan:2010gv,CMS:2012qk,Chatrchyan:2013nka,Aad:2012gla,Aad:2014lta,Abelev:2012ola,ABELEV:2013wsa}.
This anisotropy  appears in the elliptic flow coefficient $v_2$ for two particle correlations in $v_2\{2\}$, and for pA collisions in
four particle correlations $v_2\{4\}$ and for higher numbers of particles $v_2\{n\}$. 
There is also evidence for non-zero odd harmonics  of the azimuthal anisotropy,  $v_3$.  These results suggest 
that a hydrodynamic treatment may be valid~\cite{Bozek:2012gr}
in spite  of the difficulty justifying such a description for such small systems~\cite{Bzdak:2013zma}. 
It is nevertheless challenging  to reconcile with the observation that there is little evidence for jet
quenching of particles at the same transverse momentum where particles flow~\cite{CMS:2012qk}, 
and the fact that flow is seen at high transverse  momenta where hydrodynamic descriptions are questionable. %severely challenged.

The coefficient $v_2$ for two particle correlations has on the other hand been computed within 
the theory of the Color Glass Condensate~\cite{Dumitru:2008wn,Dusling:2009ni,Dumitru:2010iy,Levin:2011fb,Dusling:2012wy,Kovchegov:2012nd}, 
which for the lack of better terminology we will refer to as Glasma Graph (GG).  
This computation assumes a continuum of sources, and is in the limit of an infinite number of sources.  
The result is of order $1/N_c^2$ where $N_c$ is the number of colors. 
As was later elucidated this suppression is a manifestation of Bose-Einstein enhancement 
of gluons in the target~\cite{Altinoluk:2015uaa}, see also discussions in Ref.~\cite{Kovner:2012jm,Kovner:2011pe}. 
However, it has proven difficult to use GG mechanism to extract the four
particle correlation $v_2\{4\}$ since computations almost invariably lead
to a four particle correlation that is positive corresponding to an complex $v_2\{4\}$, see Refs.~\cite{Dumitru:2014yza,Skokov:2014tka,Dumitru:2015cfa}.

One can question the CGC treatment to date as it uses a continuum of emission sources.  In $pA$, $pp$ and peripheral $AA$ collisions, there is good evidence that the ellipticity that might drive hydrodynamic flow is generated by fluctuations in the emission from a finite number of sources~\cite{Bzdak:2013rya,Yan:2013laa,Bzdak:2013raa}.  Ellipticities have been calculated that semi-quantitatively agree with extraction from hydrodynamic models of experiment.  

In the systems driven by fluctuations, the ellipticities that fluctuations induce vanish in the limit of an infinite number of sources.
In this limit the distribution becomes uniform.
Therefore the GG CGC computations will always dominate over a fluctuation induced component for sufficiently high multiplicity in the momentum domain 
of GG CGC applicability.  
Nevertheless for the multiplicities experimentally accessible, one
might not be in the asymptotic limit.

The corresponding computation of the emission from the CGC for a finite number of sources is the subject of this paper.  
We will find that there are two effects needed to generate an acceptable
$v_2\{2\}$.  The first is the finite number of sources.  We also find that we need to have a finite range
color correlation length.  This correlation length has been computed in the literature and shown
to be of the order of the saturation momentum scale~\cite{Iancu:2002aq}.  It arises in the evolution of gluon distribution functions. 
In the McLerran-Venugopalan model~\cite{McLerran:1993ni}, the length is infinite, and one does not in this way generate an acceptable $v_2$.

Because the correlation length is of order $Q_{\rm sat}$, we will find that the characteristic momentum scale size 
associated with flow is the saturation momentum.  Again, this is a surprise since one would have naively thought 
that wave interference of the emitting gluons would be on a characteristic momentum scale of the inverse source size, 
that is $1/R_{\rm nucleon}$~\cite{Molnar:2014mwa}.

The purpose of this paper is to outline methods for the computation of $v_2\{n\}$.  We explicitly compute $v_2\{2\}$, $v_4\{2\}$ 
and $v_2\{4\}$.  We restrict our computations to transverse momentum scales greater than or of the order of that of
the nuclear saturation momentum.  There is no difficulty in principle computing in the extended range 
where $\Lambda_{\rm QCD} \ll p_\perp \ll Q_{\rm sat}$, but as a first attempt to demonstrate the method we restrict the range of momentum.

We find a semi-quantitative agreement with the experimentally measured $v_2\{2\}$.  
The magnitude and dependence upon $p_\perp$ appear to be roughly  in agreement with experiment 
for reasonable choices of our parameter.  We also
will show that the result factorizes for momentum scales consistent with out approximation 
and in accord with experimental results.  It is more complicated for us to compute $v_2\{n\}$ 
for $n \ge 4$.  We can do this as an expansion in the inverse of the number of particles sources,
and have done so for $v_2\{4\}$, and achieve an acceptable result.  The computation of $v_2\{6\}$
and higher is complicated and we do not explicitly compute here, although we point out the complications of such a computation.

We believe that computing a non-vanishing $v_3$ should be possible using the techniques we outline here.  
This involves a three gluon odderon type emission, which should generate odd two particle correlations.  
This is an interesting quantity to compute, and we hope to return to this in a later paper. 

We note that present classical Yang-Mills simulations already indicate the presence of $v_3$, it  develops 
due to  time evolution and non-linearity of the Yang-Mills equations of motion, see Ref.~\cite{Schenke:2015aqa}.

We caution the reader that the rough agreement we find with experimentally determined flow values is
probably accidental and arises from adjusting parameters, since we expect the GG  $1/N_c^2$ correction to contribute for high multiplicity events, 
and that final state interactions of particles need to be taken into account either through hydrodynamics or transport computations.
Our computation should be viewed as complementing the results
from the Glasma on initial conditions, and are amusing since these results the initial conditions have a good deal of collectivity,
in the form of initial state flow.  
Our computation supplements the results on initial state fluctuation driven ellipticity, and generate flow in the initial
state since quantum mechanics forces non-trivial
interference between the wavefunctions of particle produced from discrete sources which appear
in the momentum space amplitudes.  We also caution the reader that we are outlining here the beginnings of more detailed 
computations which allow one to extend results to wider range of momenta,
correlations with larger number of particles, and hopefully for $v_3$.

\section{Color Correlations}
We begin with a discussion of the finite correlation length for color charge in the Color Glass Condensate.  It was shown that upon evolution, there is a finite color neutralization length in the 
CGC~\cite{Iancu:2002aq}.  The specific form of the propagator was shown to have an anomalous dimension, and a range of order the inverse saturation momentum scale.  In the treatment we present here, we will include the finite range of color charge correlations but we will not include the effects of anomalous dimension.  The inclusion of an anomalous dimension presents no problem in principle but obscures the qualitative and semi-quantitative points we wish to make.

We assume that in the continuum
\begin{equation}
\langle  \rho^a(\vec{k}) \rho^b(\vec{q}) \rangle =  (2\pi)^2 \delta^{ab}\delta(\vec{k}+\vec{q}) \Delta(\vec{k}),
\label{Eq:rhorho}
\end{equation} 
where we approximate  
\begin{equation}
\Delta(\vec{k}) = \frac{\mu^2  k^2}{k^2+ m^2}.
\label{Eq:Delta}
\end{equation}
Here $m = \kappa Q_s$ and $\kappa$ is a free parameter of order one.
Equation Eq.~\eqref{Eq:Delta}  is the simplest analytic generalization of the McLerran-Venugopalan model which takes into account  
the  color neutralization phenomena discussed in details in Ref.~\cite{Iancu:2002aq}.  
It also in agreement with  the double-logarithmic approximation of BFKL 
at a very high transverse momentum.   

For future purpose we want to express $\mu^2$ in terms of the number of sources.
Lets consider the MV model 
\begin{equation}
\langle \rho^a(\vec{x})\rho^b(\vec{y}) \rangle = \delta^{ab} \delta(\vec{x}-\vec{y})  \mu^2
\label{MVrho}
\end{equation}
to compute 
\begin{equation}
\langle Q^a Q^a \rangle = d_A \int d^2x d^2y  \delta(\vec{x}-\vec{y})  \mu^2
= d_A S_\perp \mu^2,  
\label{QaQa}
\end{equation}
where $Q^a = \int d^2 x \rho^a(\vec{x})$ and $d_A=N_c^2-1$ is the dimension of
the adjoint representation. From the other hand 
\begin{equation}
\langle Q^a Q^a \rangle = {\rm tr} \left( g^2 T^a_R T^a_R N \right) 
 = g^2 C_R d_R N,  
\label{QaQafromN}
\end{equation}
where $C_R$($d_R$) is the Casimir (dimension) of the representation $R$, for the adjoint (fundamental) representation  $C_A=N_c$ ($C_F=(N_c^2-1)/2N_c$).  
Combining Eqs.~\eqref{QaQa} and \eqref{QaQafromN}, we obtain 
\begin{equation}
S_\perp \mu^2 = \frac{g^2 C_R d_R}{d_A} N. 
\label{mu2}
\end{equation}

In coordinate space,
\begin{equation}
  \Delta(\vec{x}) = \int \frac{d^2k}  {(2\pi)^2} \Delta(k) e^{i \vec{k} \vec{x}}.
\label{Eq:Deltax}
\end{equation}
The propagator $\Delta(\vec{x})$ is the charge density correlation function, i.e. 
\begin{equation}
\langle \rho^a(\vec{x}) \rho^b(\vec{y}) \rangle = \delta^{ab} \Delta(\vec{x}-\vec{y}).
\end{equation}
Explicitly,
computing the Fourier transformation of Eq.~\eqref{Eq:Deltax} we obtain 
\begin{equation}
\Delta(\vec{x}) = \mu^2 
\left[  
\delta(\vec{x}) - \frac{m^2}{2\pi} K_0(|\vec{x}| m) 
\right].
\label{Eq:DeltaX}
\end{equation}
Note that the presence of the factors of $k^2$ in the numerators of $\Delta(k)$ guarantees charge
neutralization, since then
\begin{equation}
   \int d^2x ~\Delta(x) = 0.
\end{equation}

To define $Q_s$ the scattering matrix is to be  computed  
\begin{eqnarray}
S (r_\perp) &=& \exp \left(  -g^2 C_R \int \frac{d^2k_\perp}{(2\pi)^2} \frac{\Delta(k)}{k^4} \left[1 - e^{i \vec{k}_\perp \vec{r}_\perp} \right]  \right)
= \\ 
&&\exp \left(  - \frac{g^2}{2\pi} C_R \frac{\mu^2}{m^2} \left[  \gamma_E + K_0(m r_\perp) + \ln \left( \frac{m r_\perp}{2}  \right)  \right]    \right).  
\label{S_perp}
\end{eqnarray}
Conventionally $Q_s$ is define in the limit  $m\to0$, where 
$S_\perp = \exp(- Q_{s}^2 r^2 \ln(1/\Lambda^2 r^2)  /4)$ 
with 
\begin{equation}
Q_{s}^2 = \alpha C_R \mu^2 
\label{Qs}
\end{equation}
and $\Lambda=m e^{\gamma_E-1}/2$. 
Equations~\eqref{mu2} and \eqref{Qs} give us another useful relation 
\begin{equation}
S_\perp Q_s^2 = 4 \pi  \frac{\alpha^2 C_R^2 d_R}{d_A} N. 
\label{SQs2}
\end{equation}
%In general $Q_s$ can be define by solving the equation 
%\begin{equation}
%S \left(r_\perp = \frac{\sqrt{2}}{Q_s}\right) = \exp\left(-\frac12\right) 
%\label{Sperp}
%\end{equation}
%which reads 
%\begin{equation}
%\frac{Q_{s MV}^2}{m^2}\left( \gamma_E + K_0\left(\frac{m \sqrt{2}}{Q_s} \right) + \ln\left( \frac{m\sqrt{2}}{2 Q_s} \right) \right) = \frac14.  
%\label{Qsnonlinear}
%\end{equation}

%\begin{figure}
%\centerline{\includegraphics[width=0.5\linewidth]{Qs_implicit.pdf}}
%\caption{\dots}
%\label{fig:Qs}
%\end{figure}

%The numerical solution of this equation is presented in Fig.~\ref{fig:Qs}.

Lets consider a target with sources at positions $\vec{x}_i$, $i=1,\ldots,N$.  We will assume the sources
are randomly distributed within some circular region of area $S_\perp$, where the circular region corresponds to the that of the interaction region.
Thus the color density of the target is given by 
\begin{equation}
\rho^a(\vec{x}) =\sum_i \xi_i^a \delta(\vec{x}-\vec{x}_i). 
\label{Eq:rho}
\end{equation}
The Fourier transform of the distribution is given by 
\begin{equation}
\rho^a(\vec{k}) = \int d^2 x e^{-i\vec{k}\vec{x}} \rho(\vec{x}) = \sum_i \xi_i^a e^{-i\vec{k}\vec{x}_i}.
\label{Eq:rhok}
\end{equation}
To derive the correlator for  $\xi_i^a$, lets consider 
\begin{multline}
\langle \rho^a(y) \rho^b(z) \rangle  = 
\int 
\frac{d^2x_1}{S_\perp} 
\cdots 
\frac{d^2x_N}{S_\perp}
\sum_{i,j} \langle \xi(x_i)^a \xi(x_j)^b \rangle \delta(y-x_i) \delta(z-x_j) =  \\
 \delta^{ab}  \frac{N(N-1)}{S_\perp^2} \chi(z-y)  + \delta^{ab} \frac{N}{S_\perp}
 \delta(z-y) \chi_0, 
\label{Eq:matching1}
\end{multline}
where 
\begin{equation}
\langle \xi^a(x) \xi^b(y)\rangle =  \delta^{ab} \chi(x-y)
\label{Eq:chi}
\end{equation}
and 
\begin{equation}
\langle \xi^a(x) \xi^b(x)\rangle = \delta^{ab} \chi_0.
\label{Eq:chi0prel}
\end{equation}
From Eqs.~\eqref{Eq:DeltaX} and  \eqref{Eq:chi0prel} we find 
\begin{equation}
\chi_0 = \frac{\sigma}{N},
\label{Eq:chi0}
\end{equation}
where we introduced the dimensionless transverse area with the physical meaning of the average charge squared per degree of freedom,   
\begin{equation}
\sigma = \mu^2 S_\perp. 
\label{sigma}
\end{equation}
The nontrivial dependence of the correlator on position is obtained from matching  terms of 
Eq.~\eqref{Eq:DeltaX} and  \eqref{Eq:matching1}
\begin{equation}
\chi(x) = -\frac{\sigma^2}{2\pi N(N-1)} \left( \frac{m}{\mu}  \right)^2 K_0(|\vec{x}| m).
\label{Eq:chix}
\end{equation}
The Fourier transform of the latter is 
\begin{equation}
\chi(k) = - \frac{\sigma^2}{N(N-1)} \left(  \frac{m}{\mu}   \right)^2 \frac{ 1}{k^2+ m^2}.
\label{Eq:chik}
\end{equation}

\section{S-matrix}

In this section we consider one, two and four-particle inclusive $S$-matrix. The 
results  will be used in the next section to calculated the harmonics of the azimuthal anisotropy. 

\subsection{Single particle scattering}

For fixed positions of the sources,  the S-matrix describing scattering of a quark off the target for a fixed configuration of the latter 
is 
\begin{equation}
S(\vec{y}_1,\vec{y}_2) = \frac{1}{N_c} {\rm tr} \left(  V^\dagger(\vec{y}_1) V(\vec{y}_2)  \right)
\label{Eq:S}
\end{equation}
and in the dilute limit it simplifies to 
\begin{equation}
S -1= -\frac{g^2}{4N_c} \left(\alpha^a(\vec{y}_1) - \alpha^a(\vec{y}_2)   \right)^2, 
\label{Eq:Sdl}
\end{equation}
where 
\begin{equation}
-\nabla^2 \alpha^a(\vec{x}) = \rho_a(\vec{x})
\label{Eq:Poisson}
\end{equation}
and thus 
\begin{equation}
\alpha^a(\vec{k}) = \sum_i \xi_i^a \frac{1}{k^2} e^{-i\vec{k}\vec{x}_i} .
\label{Eq:alphak}
\end{equation}

Performing Fourier transformation we obtain 
\begin{equation}
S(\vec{p}, \vec{b})-1  =- \frac{g^2}{N_c} \int \frac{d^2k}{(2\pi)^2} 
\left[ \alpha^a(\vec{k}+\vec{p})\alpha^a(\vec{p}-\vec{k}) e^{2 i \vec{b}\vec{p}}  
+ {\rm c.c.}
-2 \alpha^a(\vec{k}+\vec{p})\alpha^a(\vec{k}-\vec{p}) e^{2i\vec{k}\vec{b}}
\right].
\label{Eq:S_FT}
\end{equation}
Integrating with respect to the impact parameter one gets
%\begin{equation}
%S = - \frac{g^2}{2N_c} \alpha^a(\vec{y}_1) \alpha^a(\vec{y}_2)
%\label{Eq:Sdlsim}
%\end{equation}
%or in the momentum space
\begin{equation}
S(\vec{p}, \vec{x}_1, \ldots, \vec{x}_N) -1 = - \frac{g^2}{2N_c}
\left( \delta(\vec{p}) \int d^2 k |\alpha^a(k)|^2 - |\alpha^a(\vec{p})|^2
\right).
\label{Eq:Sp}
\end{equation}
The first term is not of the interest to us; it is related to the inverse is of the projectile. 
Substituting  Eq.~\eqref{Eq:alphak} we get 
\begin{equation}
S(\vec{p}, \vec{x}_1, \ldots, \vec{x}_N) -1= \frac{g^2}{2N_c}\frac{1}{p^4} \sum_{i,j} \xi_i^a \xi_j^a e^{-i\vec{p}(\vec{x}_i-\vec{x}_j)} = 
\frac{g^2}{2N_c} \frac{1}{p^4} \left(  \xi_i^a \xi_i^a  + \sum_{i\ne j}  \xi_i^a \xi_j^a e^{-i\vec{p}(\vec{x}_i-\vec{x}_j)}     \right).
\label{Eq:Spp}
\end{equation}
The experimentally  measured  one particle inclusive S-matrix can be obtained by averaging Eq.~\eqref{Eq:Spp} with respect to all possible sources' positions 
and configurations of $\xi_i^a$. 
\subsection{Two particle scattering}
To extract the two-particle azimuthal anisotropy, the two particle S-matrix is to be computed 
\begin{multline}
\langle S_2(\vec{p}_1, \vec{p}_2) \rangle = S_\perp^{-N} \int \left( \Pi_{l}  d^2 x_l \right) \langle S(\vec{p}_1, \vec{x}_1, \ldots, \vec{x}_N ) S(\vec{p}_2, \vec{x}_1, \ldots, \vec{x}_N )  \rangle =\\ 
S_\perp^{-N} \int \left( \Pi_{l}  d^2 x_l \right)
\sum_{i,j}
\sum_{i',j'}
\left(\frac{g^2}{2N_c}\right)^2 \frac{1}{p_1^4 p_2^4}
\left\langle 
\sum_{i,j} \xi_i^a \xi_j^a e^{-i\vec{p}_1(\vec{x}_i-\vec{x}_j)} 
\sum_{i',j'} \xi_{i'}^b \xi_{j'}^b e^{-i\vec{p}_2(\vec{x}_{i'}-\vec{x}_{j'})} 
\right\rangle \ \ \ \ \ 
\myeq \\
\left(\frac{g^2}{2N_c}\right)^2 \frac{(N_c^2-1)^2}{p_1^4 p_2^4} \frac{1}{S_\perp^2} \left( \chi(p_1) \chi(p_2) \left[ \frac{N!}{(N-4)!} + 4\frac{N!}{(N-3)!} \right] + 
\frac{N N! S_\perp}{(N-2)!} \chi_0 \left[\chi(p_1)+\chi(p_2)\right]+\right.\\\left.
S_\perp^2 N^2  \chi_0^2 + \frac{S_\perp N!}{(N-2)!} \left[ \chi_2(p_1+p_2)+\chi_2(p_1-p_2)\right] 
\right), 
\label{Eq:S2}
\end{multline}
where 
\begin{equation}
\chi_2(p) = \int d^2 x e^{ipx} \chi^2(x) = \frac{2 \pi}{m^2} \left[\frac{\sigma^2}{ 2\pi N(N-1)}\left(\frac{m}{\mu}  \right)^2  \right]^2  I(p), \ \ \ I(p)=\frac{ m\  {\sinh^{-1}}\left(\frac{p}{2m}\right)}{p \sqrt{1+\left(\frac{p}{2m}\right)^2}}. 
\label{Eq:chi2}
\end{equation}
For small (large)   $p$, $I(p\ll m ) \approx  \frac12 \left(1-\frac{p^2}{6m^2}\right)$  ($I(p\gg m) \approx  \frac{m^2}{p^2} \ln\left(\frac{p^2}{m^2}\right)$).
The second equality in Eq.~\eqref{Eq:S2} is only valid in the large $N_c$ limit because the connected contributions (in term of the  color indices),
also called the ``glasma'' graphs
were  neglected. They are however  important to be accounted for in the limit of infinite number of sources, as we discussed in the Introduction. 

The azimuthal anisotropy can arise only from the last term of Eq.~\eqref{Eq:S2}.
The Fourier components of $\chi_2(p_1+p_2)$ for an even $n$ are given by 
\begin{multline}
\langle \chi_2(p_1+p_2) e^{i n (\phi_1-\phi_2)}\rangle \equiv
\int \frac{d\phi_1}{2\pi} \frac{d\phi_2}{2\pi} \chi_2(p_1+p_2) = 
\int d^2 x J_n(|p_1| |x|) J_n(|p_2| |x|) \chi^2(x) = \\2\pi \int d x  x J_n(|p_1| |x|) J_n(|p_2| |x|) \chi^2(x). 
\label{Eq:Fchi}
\end{multline}
An analytic expression for the Fourier components can be derived 
if $|p_1|=|p_2|=p$
\begin{equation}
\langle \chi_2(p_1+p_2) e^{i n (\phi_1-\phi_2)}\rangle_{|p_1|=|p_2|=p} 
= \frac{2 \pi}{m^2} \left[\frac{\sigma^2}{ 2\pi N(N-1)}\left(\frac{m}{\mu}  \right)^2  \right]^2  I_n  (p).
\label{Eq:Fchisp}
\end{equation}
For the relevant values of $n$ we get   
\begin{equation}
I_0(p) = \frac{m} {2p}\tan^{-1} \left( \frac{p}{m}\right)
\label{Eq:I0}
\end{equation}
and 
\begin{equation}
I_2(p) = I_0(p)+ \frac{m^2}{p^2} \left[ 1 -  \left(1+\frac{m^2}{p^2}\right)\ln\left(1+\frac{p^2}{m^2}\right) \right].
\label{Eq:I2 }
\end{equation}

In the next section  the numerical computations of Eq.~\eqref{Eq:S2} will be performed, here we only want 
to discuss the general structure of the S-matrix. To simplify matters for now, let us consider the large $N$ limit 
of Eq.~\eqref{Eq:S2}: 
\begin{multline}
\langle S_2(\vec{p}_1, \vec{p}_2) \rangle_{ {\rm Large }\  N } =
\left(\frac{g^2}{2N_c}\right)^2 \frac{(N_c^2-1)^2}{p_1^4 p_2^4} \sigma^2
  \left[ 
  \frac{1}{(p_1^2+m^2)(p_2^2+m^2)}\left( p_1^2 p_2^2 - \frac{2m^4}{N^2}\right)
  +\right.\\ \left. \frac{\sigma}{2\pi N^2}\left(\frac{m}{\mu}\right)^2  \left( I(\vec{p}_1+\vec{p}_2) +I(\vec{p}_1-\vec{p}_2) \right)
\right].  
\label{Eq:S2largeN}
\end{multline}
As can be seen from Eq.~\eqref{Eq:S2largeN} the azimuthally  anisotropic part is proportional to $\frac{\sigma}{N^2} \left( \frac{m}{\mu} \right)^2$. 
Previously we established  that  $\sigma \propto N $ and thus the amplitude of the azimuthal anisotropy is suppressed by 
$N^{-1}$ instead of a naive $N^{-2}$. The remaining factor of $\left( \frac{m}{\mu} \right)^2 = \alpha C_R \kappa^2$. Thus the amplitude is also controlled 
by the free parameter $\kappa$. The factor of $\alpha C_R$ is of order one in the relevant energy range.       

\subsection{Four particle scattering}
Due to its phenomenological importance we also compute four-particle S-matrix. We were able to do this only in 
the limit $|p_1|=|p_2|=|p_3|=|p_4|=p$ and we neglected the contributions suppressed  by powers of $N$ 
when they are not compensated by large combinatorics factors. The full calculations of the four-particle S-matrix 
become very involved and will not make any significant difference for the numerical results presented in the next section.
Straightforward computations yield  
\begin{multline}
\langle S_4(\vec{p}_1, \vec{p}_2, \vec{p}_3,\vec{p}_4) \rangle =
\left(\frac{g^2}{2N_c}\right)^4 \left(\frac{N_c^2-1}{p^4} \right)^4
\\  \left\{
\frac{\chi^4(p)}{S_\perp^4}
\left[
\frac{N!}{(N-8)!}
+2^2 \binom 24  \frac{N!}{(N-7)!}
+ 2^4 \binom 24  \frac{N!}{(N-6)!}
+ 2^3 \binom 14 \binom 13  \frac{N!}{(N-6)!}
+ 2^5 \binom 24   \frac{N!}{(N-5)!}
\right] + \right. \\ 
\frac{\chi^3(p)}{S_\perp^3} \binom 14 N  \chi_0  
\left[
 \frac{N!}{(N-6)!} +  
 2^2 \binom 13  \frac{N!}{(N-5)!} +  
 2^3 \binom 13  \frac{N!}{(N-4)!} 
\right] + \\ 
 \chi_0^2 \frac{\chi^2(p)}{S_\perp^2}
 \binom24 N^2 
 \left[
 \frac{N!}{(N-4)!}
 + 2^2  \frac{N!}{(N-3)!}
\right]  
+  \chi_0^3 \frac{\chi(p)}{S_\perp}
  \binom 14  N^3 \frac{N!}{(N-2)!}+ N^4 \chi_0^4+ \\
%24 \frac{N!}{(N-7)!} \frac{\chi^4(p)}{S_\perp^4}
%+48  N \frac{N!}{(N-5)!} \chi_0 \frac{\chi^3(p)}{S_\perp^3}
%+24  N^2 \frac{N!}{(N-3)!} \chi_0^2 \frac{\chi^2(p)}{S_\perp^2} +
%\\
\left. 
\frac{N!}{(N-4)!} 
\left[
\left( \frac{\chi_2(p_1+p_2)}{S_\perp} + \frac{\chi_2(p_1-p_2)}{S_\perp} \right)
\left( \frac{\chi_2(p_3+p_4)}{S_\perp} + \frac{\chi_2(p_3-p_4)}{S_\perp} \right)+\right. \right.\\
\left( \frac{\chi_2(p_1+p_3)}{S_\perp} + \frac{\chi_2(p_1-p_3)}{S_\perp} \right)
\left( \frac{\chi_2(p_2+p_4)}{S_\perp} + \frac{\chi_2(p_2-p_4)}{S_\perp} \right)+\\
\left.\left.
\left( \frac{\chi_2(p_1+p_4)}{S_\perp} + \frac{\chi_2(p_1-p_4)}{S_\perp} \right)
\left( \frac{\chi_2(p_2+p_3)}{S_\perp} + \frac{\chi_2(p_2-p_3)}{S_\perp} \right)
\right] + 
\right.
\\ 
\left. \frac{N!}{(N-2)} 
\left[ 
\frac{\chi_4(p_1+p_2+p_3+p_4)}{S_\perp} + \ \ {\rm sign\ permutations\ in\ front\ of\ } p_{2,3,4}
\right]
  \right\} = 
\left(\frac{g^2}{2N_c}\right)^4 \left(\frac{N_c^2-1}{p^4} \right)^4
\\  \left\{
\frac{\chi^4(p)}{S_\perp^4}
\left[
\frac{N!}{(N-5)!} (N^3+6N^2+35N-258)
\right] + \right. \\ 
\frac{\chi^3(p)}{S_\perp^3} 4 N  \chi_0  
\left[
 \frac{N!}{(N-4)!}(N-1)(N+4)  
\right] + \\ 
 \chi_0^2 \frac{\chi^2(p)}{S_\perp^2}
 6 N^2 
 \left[
 \frac{N!}{(N-3)!} (N+1)
\right]  
+  \chi_0^3 \frac{\chi(p)}{S_\perp}
  4 N^4 (N-1)+ N^4 \chi_0^4+ \\
%24 \frac{N!}{(N-7)!} \frac{\chi^4(p)}{S_\perp^4}
%+48  N \frac{N!}{(N-5)!} \chi_0 \frac{\chi^3(p)}{S_\perp^3}
%+24  N^2 \frac{N!}{(N-3)!} \chi_0^2 \frac{\chi^2(p)}{S_\perp^2} +
%\\
\left. 
\frac{N!}{(N-4)!} 
\left[
\left( \frac{\chi_2(p_1+p_2)}{S_\perp} + \frac{\chi_2(p_1-p_2)}{S_\perp} \right)
\left( \frac{\chi_2(p_3+p_4)}{S_\perp} + \frac{\chi_2(p_3-p_4)}{S_\perp} \right)+\right. \right.\\
\left( \frac{\chi_2(p_1+p_3)}{S_\perp} + \frac{\chi_2(p_1-p_3)}{S_\perp} \right)
\left( \frac{\chi_2(p_2+p_4)}{S_\perp} + \frac{\chi_2(p_2-p_4)}{S_\perp} \right)+\\
\left.\left.
\left( \frac{\chi_2(p_1+p_4)}{S_\perp} + \frac{\chi_2(p_1-p_4)}{S_\perp} \right)
\left( \frac{\chi_2(p_2+p_3)}{S_\perp} + \frac{\chi_2(p_2-p_3)}{S_\perp} \right)
\right] 
  \right\} +\\ 
\frac{N!}{(N-2)} 
\left[ 
\frac{\chi_4(p_1+p_2+p_3+p_4)}{S_\perp} + \ \ {\rm sign\ permutations\ in\ front\ of\ } p_{2,3,4}
\right].
\end{multline}
Here we introduced  
\begin{equation}
\chi_4 (p) = \int d^2 x e^{i x p} \chi^4(x). 
\label{chi4}
\end{equation}
For the numerical calculations presented in the next section we  need only the following  angular average 
\begin{multline}
\langle \chi_4 (p_1\pm p_2\pm p_3\pm p_4)  e^{i n (\phi_1 + \phi_2 - \phi_3 -\phi_4)} \rangle_{|p_i|=p} 
= \\
\frac{2\pi}{m^2}  \left[\frac{\sigma^2}{ 2\pi N(N-1)}\left(\frac{m}{\mu}  \right)^2  \right]^4  Y_n(p), \ \ Y_n(p) = \int_0^\infty dx x J_n^4\left(\frac{p}{m} x\right) K_0^4(x). 
\label{chi4ang}
\end{multline}

\section{Harmonics of azimuthal anisotropy}
Using the results of the previous section we will extract the harmonics of the azimuthal anisotropy. 
The derivation was performed for the fundamental representation of the Wilson lines, but in the dilute limit, 
we operate in, the Casimirs and the dimension of representations enter as  overall numerical constants. Thus 
the results in the adjoint representation for the observables we are about to consider will be the same as  for the
fundamental. 

\begin{figure}
\centerline{\includegraphics[width=0.7\linewidth]{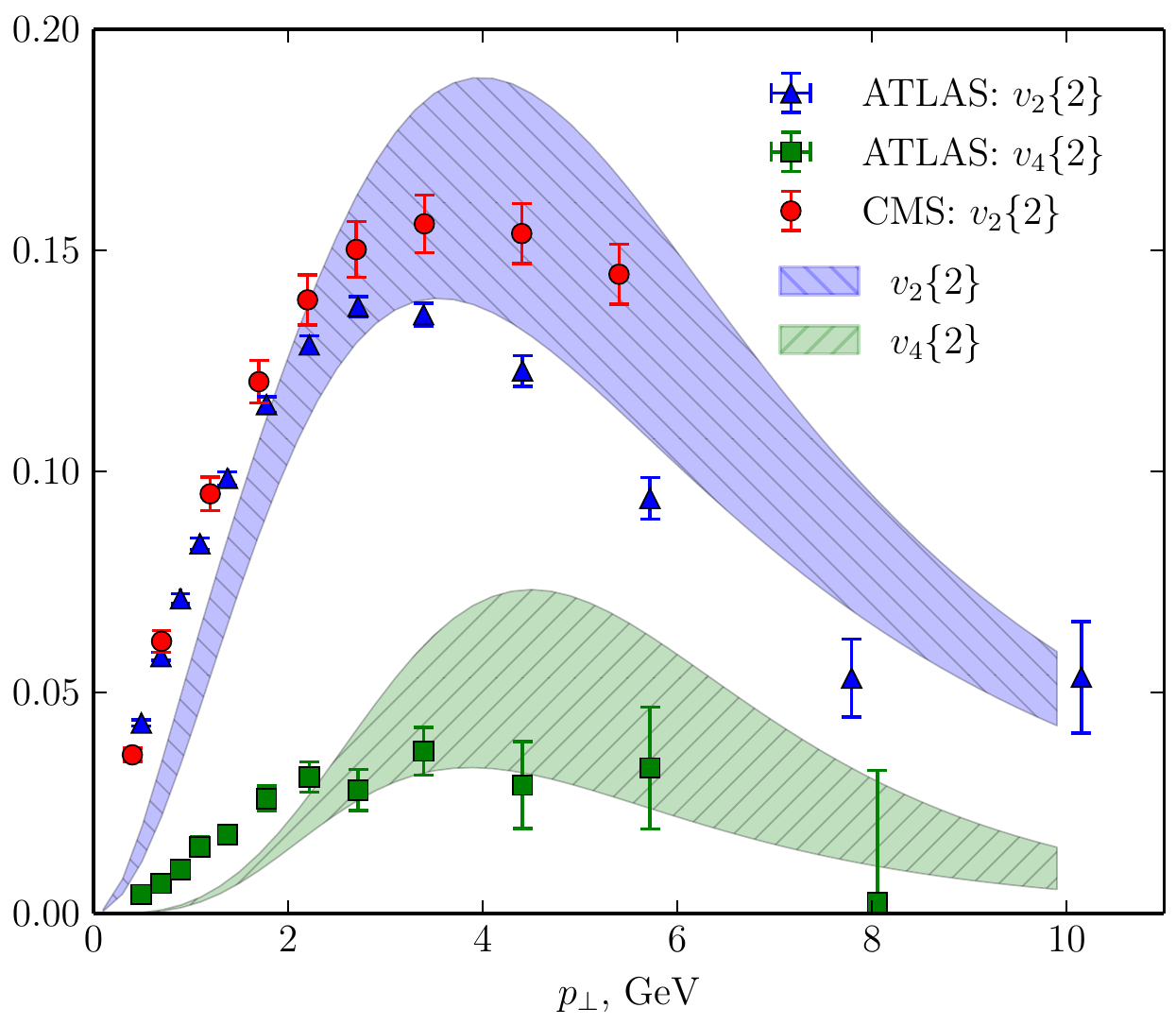}}
\caption{The azimuthal anisotropy harmonics $v_2\{2\}$ and $v_4\{2\}$ as a function of the transverse momentum.  
The bands correspond to the model results with $p_\perp^{\rm Ref}$ varying from 1 to 3 GeV. The ATLAS data is for 1 GeV $<p_\perp<$ 3 GeV and 
$220<N_{\rm ch}<260$, see Ref.~\cite{Aad:2014lta}.
 The CMS data is for $0.3$ GeV $<p_\perp<$ 3 GeV and 
$220<N_{\rm ch}<260$, see Ref.~\cite{Chatrchyan:2013nka}. The model parameters are $Q_s=1.2$ GeV, $\kappa=1.7$ and $N=15$.  }
\label{fig:v22}
\end{figure}

For two-particle azimuthal anisotropy, we follow the definitions commonly accepted in the community. The two-particle azimuthal anisotropy is given by 
\begin{equation}
v_2\{n\} (p_\perp) = \frac{V_{n\Delta}(p_\perp, p_\perp^{\rm Ref})}{\sqrt{V_{n\Delta}(p_\perp^{\rm Ref}, p_\perp^{\rm Ref}))}},  
\label{Eq:v2ndef}
\end{equation}
where 
\begin{equation}
V_{n\Delta}(p_\perp^{\rm A}, p_\perp^{\rm B})  = \frac{ \langle S_2(p_\perp^{\rm A}, p_\perp^{\rm B}) \exp \left( i n \Delta \phi \right)   \rangle } { \langle S_2(p_\perp^{\rm A}, p_\perp^{\rm B})  \rangle }. 
\label{Eg:VnD}
\end{equation}
In our model we have a few free parameters, $Q_s$, $\kappa$ and $N$. The number of sources can be in principal fixed by performing Monte-Carlo simulations 
and triggering number of produced particles. This however will require an access to the soft momentum range, where the dilute approximation breaks down and 
our model cannot be formally applied. Thus we keep the number of sources to be a free parameter ranging from 10 to 25. The results of the fit of the experimental
data  results in $N=15$. Again we present this result with a grain of salt: given the approximations we made they are at best qualitative and cannot be considered as 
a quantitative outcome of the model. Nonetheless we think it is important to demonstrate the model ability to fit the experimental data with a reasonable parameter set. 
The results for this fit are presented in Fig. 1. We used  $Q_s$, $\kappa$ and $N$ to fit $v_2\{2\}$; $v_4\{2\}$ is the ``prediction'' of the model.  

\begin{figure}
\centerline{\includegraphics[width=0.7\linewidth]{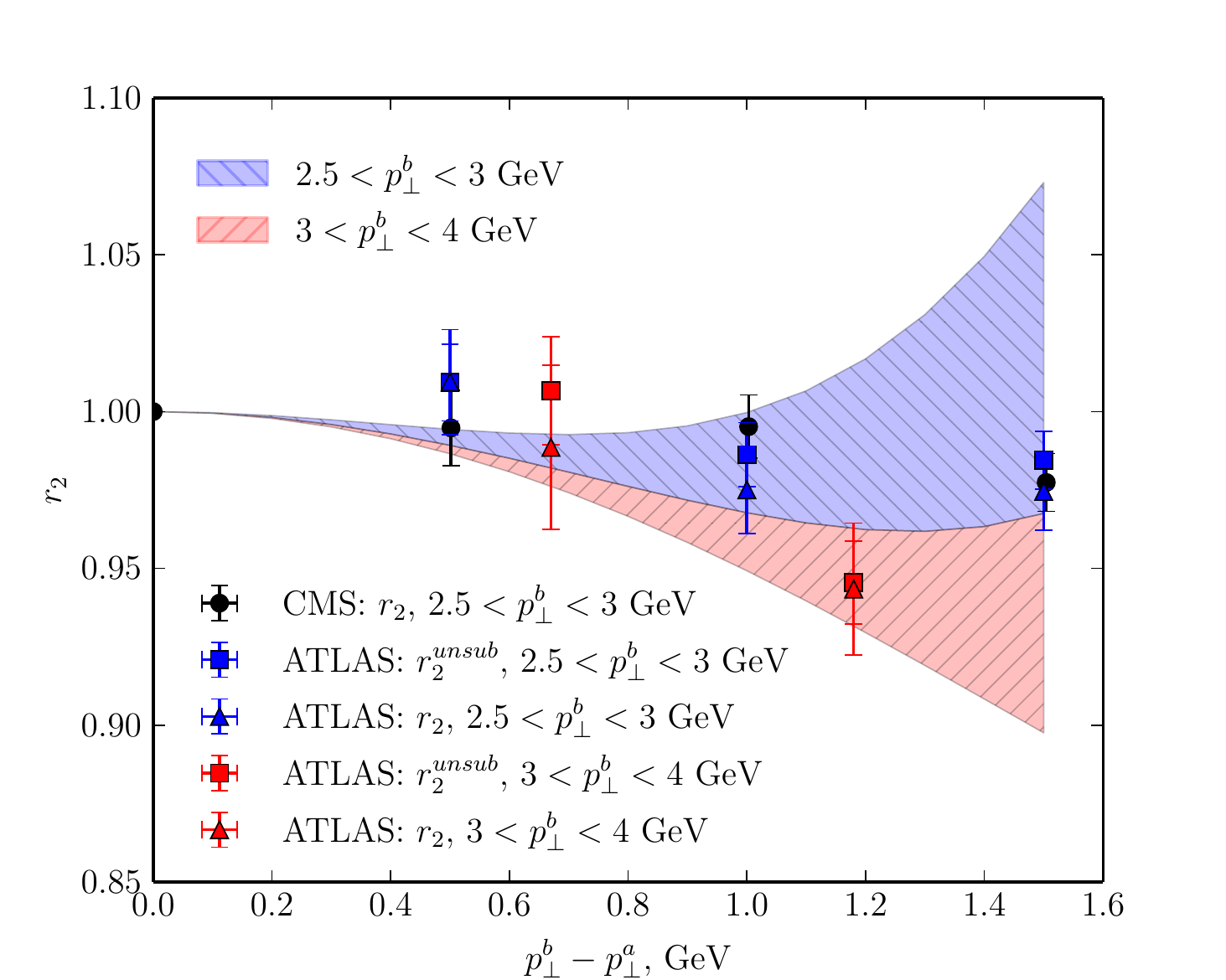}}
\caption{The ratio $r_2$ at a function of the momentum difference. The ATLAS and CMS data are from Refs.~\cite{Aad:2014lta,Khachatryan:2015oea} respectively. 
The experimental data are for $220<N_{\rm ch}<260$. The model parameters are the same as in Fig.~\ref{fig:v22}.}
\label{fig:r2}
\end{figure}

Lets also discuss another important observable which probes  decorrelation of the harmonics in  the transverse momentum. It  is defined by  
\begin{equation}
r_n = \frac{V_{N\Delta}(p_\perp^{\rm A}, p_\perp^{\rm B}) }{  \sqrt{V_{N\Delta}(p_\perp^{\rm A},  p_\perp^{\rm A}) V_{N\Delta}(p_\perp^{\rm B}, p_\perp^{\rm B}) } }. 
\label{Eq:rn}
\end{equation}
In the model the decorrelation is governed by $m^2$ and thus is close to one and does not depend on momentum at very high momenta. 
In the range of the model applicability we obtained the results shown in the Fig. 2. The description of the data is satisfactory.

\begin{figure}
\centerline{\includegraphics[width=0.7\linewidth]{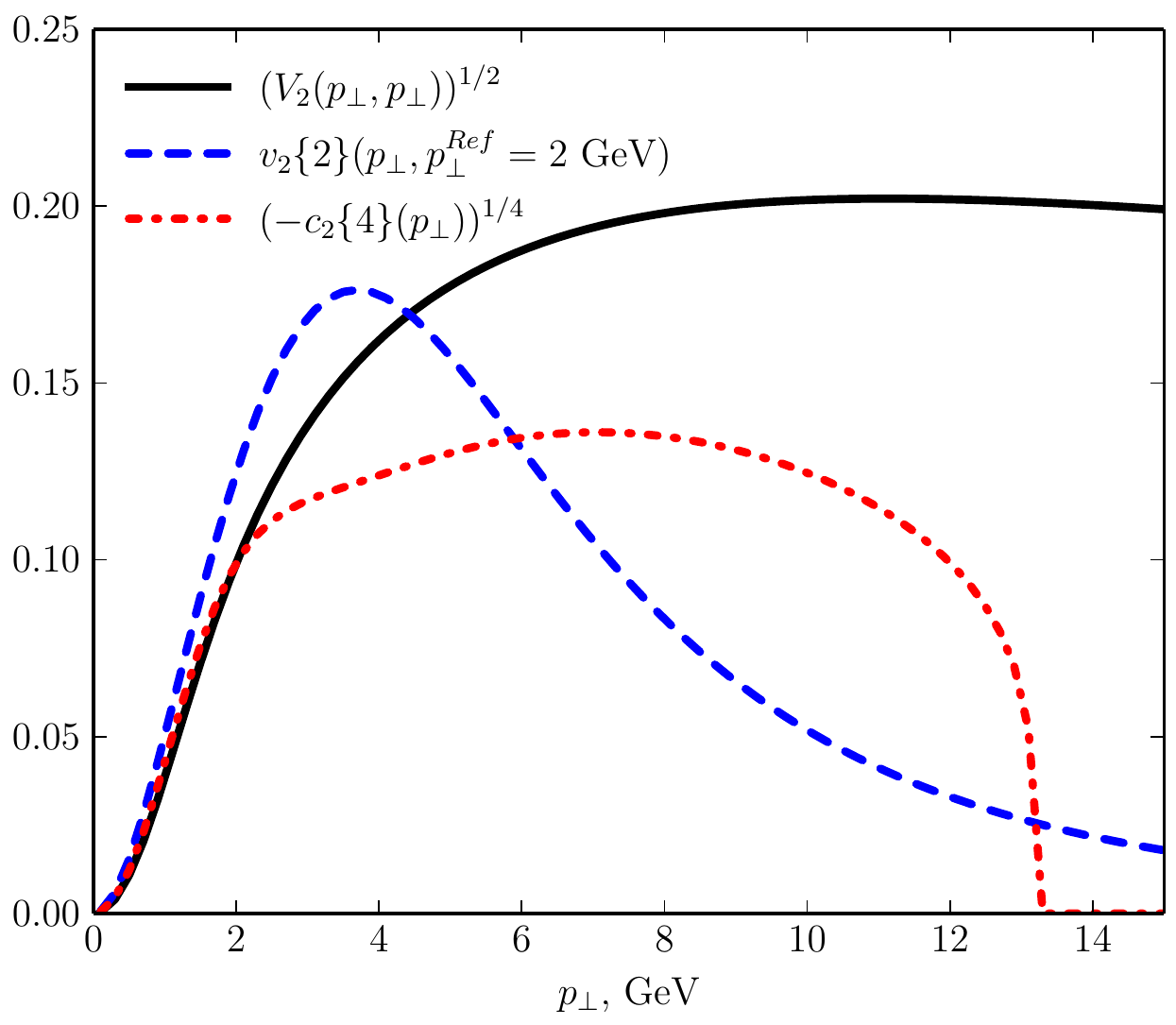}}
\caption{
Comparison of the second Fourier harmonics computed from two-particle correlation function 
at the same momentum $(V_{2\Delta})^{1/2}$ and using the definition~\eqref{Eq:v2ndef}.    
}
\label{fig:v2comp}
\end{figure}

Since we were not able to compute the four-particle correlation function for arbitrary momenta, the direct comparison 
with the experimental data for  $v_2\{4\}$ is not possible. Nevertheless, to qualitatively examine the four-particle azimuthal anisotropy 
we compute $c_2\{4\}(p_\perp)$ when all particles are taken at the same momenta. The result for $N=10$,
$Q_s=1.2$ GeV and  $\kappa=1.7$ 
is presented in Fig.~\ref{fig:v2comp}.

We compare  $(-c_2\{4\}(p_\perp))^{1/4}$ to $v_2\{2\}$ and $(V_{n\Delta}(p_\perp,p_\perp))^{1/2}$.   
First of all there is a very striking difference between two different ways to compute the two-particle azimuthal 
anisotropy: the $v_2\{2\}$, computed according to the experimental prescription Eq.~\eqref{Eq:v2ndef},  decays fast with the momentum, while  $(V_{n\Delta}(p_\perp,p_\perp))^{1/2}$ show almost no dependence on the momentum at 
high $p_\perp$. Second, the four particle azimuthal anisotropy defined by   $(-c_2\{4\}(p_\perp))^{1/4}$ is very close to the one defined by two particles. 
Moreover $c_2\{4\}(p_\perp))$ is negative in the range of momentum relevant to the phenomenology and changes sign at about $13.5$ GeV.  
Note that this numerical value may change once the connected contributions of  GG or other local in rapidity  background effects are taken into account, which presumable will lower this number.

\section{Conclusions}
In this paper we performed first computations of the azimuthal anisotropy in CGC with finite number of sources for momentum of particles greater than or of the order of the saturation momentum.   This computation should be viewed as the precursor to refinements which allow computation for $\Lambda_{QCD} \ll p_T \ll Q_{sat}$, and for higher order particle cumulants such as $v_2\{n\}$ for $n \ge 6$.  We also need to find a way to compute an initial $v_3\{2\}$.

This computation should be viewed in a larger framework:  What are the initial values of ellipticities and flow moment predicted by first principles in QCD?  Then another question is how are these initial values modified by final state interactions, be they hydrodynamic or transport generated.
Without proper first principle computation of both final state and initial state physics, the beautiful experimental results showing flow in 
the high multiplicity collisions of small systems cannot be
properly understood.

\section{Acknowledgements}

The authors gratefully acknowledge the comments of Jean Paul Blaizot, Soren Schlichting, Bjoern Schenke, Raju Venugopalan and Ulrich Heinz.

Larry McLerran is supported under Department of Energy contract number Contract No. DE-SC0012704.

\end{document}